\documentclass[pre, letterpaper, twocolumn, fleqn, floatfix, showpacs, showkeys ]{revtex4}

\usepackage{graphicx}% Include figure files
\usepackage{dcolumn}% Align table columns on decimal point
\usepackage{bm}% bold math
\usepackage[citecolor=blue]{hyperref}

\newcommand{\wtV}{\widetilde{V}}
\newcommand{\wtn}{\widetilde{n}}
\newcommand{\wtN}{\widetilde{N}}
\newcommand{\wtsig}{\widetilde{\sigma}}
\newcommand{\wtC}{\widetilde{C}}

\begin{document}

\preprint{APS/123-QED}

\title{Modeling electrolytically top gated graphene}
\author{Z.\ L.\ Mi\v{s}kovi\'{c}}
\affiliation{Department of Applied Mathematics, University of Waterloo, Waterloo, Ontario, Canada N2L 3G1}
\author{Nitin Upadhyaya}
\affiliation{Department of Applied Mathematics, University of Waterloo, Waterloo, Ontario, Canada N2L 3G1}

\date{\today}

\begin{abstract}

We investigate doping of a single-layer graphene in the presence of electrolytic top gating. The interfacial phenomena is
modeled using a modified Poisson-Boltzmann equation for an aqueous solution of simple salt. We demonstrate both the sensitivity
of graphene's doping levels to the salt concentration and the importance of quantum capacitance that arises due to the smallness
of the Debye screening length in the electrolyte.

\end{abstract}

 \maketitle

\section{Introduction}

Carbon nano-structures show great promise in many applications, including chemical and biological sensors. While carbon
nanotubes (CNTs) have been extensively studied in that context for quite some time \cite{Kong_2000,Kauffman_2008},
investigations of graphene as a sensor are only beginning to appear \cite{Schedin_2007, Robinson_2008, Ang_2008}. Sensory
function of carbon nano-structures is generally implemented in the configuration of a field effect transistor (FET), with a
prominent role played by the gate potential that controls the current through the device. Biochemical applications require good
understanding of the interaction of carbon nano-structures with aqueous solutions \cite{Ang_2008}, often in the context of the
electrochemical top gating \cite{Das_2008}. While significant progress has been achieved in understanding the interaction of
CNT-FETs with the electrolytic environment \cite{Kruger_2001,Rosenblut_2002,Heller_2006}, similar studies involving graphene
have appeared only very recently \cite{Das_2008}, focusing on the screening effect of an ion solution on charge transport
through graphene based FETs \cite{Chen_2009}, as well as on the measurement of the quantum capacitance of graphene as an
ultimately thin electrode in an aqueous solution \cite{Xia_2009}.

The top gating of a graphene based FET with a solid or liquid electrolyte presents several advantages compared to the
conventional back gating with a metallic electrode. Upon application of gate voltage, free ions in the electrolyte re-distribute
themselves, forming an electrostatic double layer (EDL) at the interface between graphene and the electrolytic solution
\cite{Castro_2009}. Depending on the ion concentration, the EDL can be only a few nanometers thick, while still providing
efficient shielding of graphene. As a consequence, the capacitance of the EDL in an electrolyte can be much higher than the
capacitance of the back gate, which is typically separated from graphene with a layer of SiO$_2$ a few hundred nanometers thick
\cite{Castro_2009}. This property of the EDL enables a much better control of the surface potential on the graphene layer, while
requiring a much lower operating voltage that needs to be applied to the reference electrode in the electrolyte than voltages
currently used with back gates \cite{Castro_2009}. The applied voltage then modifies the chemical potential of graphene,
resulting in changes in its observable properties such as conductance. Since properties of the EDL depend on the ion
concentration, monitoring the resulting changes in graphene's conductance can provide a means for sensor application, e.g., in
measuring the amount of salt in the solution.

On the other hand, referring to the electrical model of the electrolytic gating as a series connection of capacitors
\cite{Bard_2001}, the high gate capacitance in the electrolyte gives a much more prominent role to the quantum capacitance of
graphene than does the back gate \cite{Xia_2009,Fang_2007,Guo_2007,Giannazzo_2009}. In addition, doping levels of an
electrolytically top gated graphene have been reported recently \cite{Das_2008} to be much higher than those obtained with the
conventional back gate \cite{Sonde_2009}. At the same time, mobile ions in the solution seem to provide a much more effective
screening of charged impurities underneath the graphene, thereby significantly increasing the charge carrier mobility in
graphene in comparison to some other high-$\kappa$ dielectric environments \cite{Ponomarenko_2009}. All these facts indicate
that electrolytic top gating provides a means to develop high performance FETs.

While the above few experimental observations reveal quite fascinating aspects of the graphene-electrolyte interaction,
theoretical modeling of this system seems to be lagging behind the experiment. It is therefore desirable, and tempting to
discuss doping of a single layer of graphene by a remote gate electrode immersed in a thick layer of electrolyte by using two
simple models: one describing graphene's $\pi$ electron band structure in the linear energy dispersion approximation
\cite{Castro_2009}, and the other describing the distribution of ions in the electrolyte by a one-dimensional (1D)
Poisson-Boltzmann (PB) model, which takes advantage of the planar symmetry of the problem \cite{Bard_2001}. However, it should
be emphasized that the experiments involving electrolytic top-gating of both carbon nanotubes \cite{Kruger_2001,Rosenblut_2002}
and graphene \cite{Das_2008,Chen_2009} use rather hight voltages, on the order of 1-2 V, which can cause significant crowding of
counter-ions at the electrolyte-graphene interface. It is therefore necessary to go beyond the standard PB model by taking into
account the steric effects, i.e., the effects of finite size of ions in the solution. To that aim, we shall use the modified PB
(mPB) model developed by Borukhov \emph{et al.}\cite{Borukhov_1997,Kilic_2007}, which retains analytical tractability of the
original 1D-PB model. In addition, applied voltages beyond 1 V also require taking into account non-linearity of graphene's band
energy dispersion, giving small but noticeable corrections to the linear approximation.

We shall consider here a simple 1:1 electrolyte representing an aqueous solution of NaF because both the Na$^+$ and F$^-$ ions
are chemically inert allowing us to neglect their specific adsorption on the graphene surface \cite{Chen_2009,Bard_2001}. In
particular, we shall analyze the density of doped charge carriers in graphene at room temperature (RT) as a function of both the
applied voltage and the salt concentration to elucidate graphene's sensor ability. In addition, we shall evaluate the
contributions of both graphene and the EDL in the total gate capacitance in terms of the applied voltage to reveal the
significance of quantum capacitance, as well as to elucidate the behavior of the EDL under high voltages. We shall cover broad
ranges of both the salt concentration, going from $\mu$M to a physiologically relevant value, and the applied voltage, going up
to about 2 V.

After outlining in section II our theoretical models for graphene and the EDL layer, we shall introduce several reduced
quantities of relevance for these two vastly different systems and present our results in section III. Concluding remarks are
given in section IV. Note that we shall use gaussian units ($4\pi\epsilon_0=1$) throughout the paper, unless otherwise
explicitly stated.

\section{Theoretical model}

Graphene is a semi-metal, or a zero-gap semiconductor because its conducting and valence $\pi$ electron bands touch each other
only at two isolated points in its two-dimensional (2D) Brillouin zone \cite{Castro_2009}. The conical shape of these bands in
the vicinity of these points gives rise to an approximately linear density of states,
$\rho_L(\varepsilon)=g_d\vert\varepsilon\vert/[2\pi(\hbar v_F)^2]$, where $g_d=4$ is the spin and the band valley degeneracy
factor, and $v_F\approx c/300$ is the Fermi speed of graphene, with $c$ being the speed of light in vacuum \cite{Castro_2009}.
In the intrinsic, or undoped graphene, the Fermi energy level sits precisely at the neutrality point, $\varepsilon_F=0$, also
called the Dirac point. Therefore, the electrical conductivity of graphene is easily controlled, e.g., by applying a gate
voltage $V_A$ that will cause doping of graphene's $\pi$ bands with electrons or holes (depending on the sign of $V_A$), which
can attain the number density per unit area, $n$, with a typical range of $n\sim 10^{11}$ to $10^{13}$ cm$^{-2}$
\cite{Castro_2009}. In a doped graphene Fermi level moves to $\varepsilon_F=\hbar v_F\sqrt{\pi\vert n\vert}\,\mathrm{sgn}(n)$,
where $\mathrm{sgn}(n)=\pm 1$ for electron (hole) doping. At a finite temperature $T$, one can express the charge carrier
density in a doped graphene in terms of its chemical potential $\mu$ as \cite{Radovic_2008}
\begin{eqnarray}
n(\mu)=\int\limits_0^\infty d\varepsilon\,\rho(\varepsilon)\left[\frac{1}{1+\mathrm{e}^{\beta(\varepsilon-\mu)}}-\frac{1}{
1+\mathrm{e}^{\beta(\varepsilon+\mu)}}\right], \label{density}
\end{eqnarray}
where $\beta\equiv \left(k_BT\right)^{-1}$ with $k_B$ being the Boltzmann constant. We shall use in our calculations a full,
non-linear expression for the $\pi$ electron band density, $\rho(\varepsilon)$, given in Eq.\ (14) of Ref.\cite{Castro_2009}.
However, for the sake of transparency, the theoretical model for graphene will be outlined below within the linear density
approximation, $\rho(\varepsilon)\approx\rho_L(\varepsilon)$. We note that this approximation is accurate enough for low to
moderate doping levels, such that, e.g., $\vert\mu\vert\lesssim 1$ eV, and it only incurs a relative error of up to a few
percent when $1\lesssim\vert\mu\vert\lesssim2$ eV.

At this point, it is convenient to define the potential $V_Q=-\mu/e$, where $e>0$ is the proton charge, which is associated with
the quantum-mechanical effects of graphene's band structure \cite{Rossier_2007}, and relate it to the induced charge density per
unit area on doped graphene, $\sigma=-en$, via the Eq.\ (\ref{density}),
\begin{eqnarray}
\sigma=\frac{2}{\pi}\frac{e}{\left(\hbar v_F\beta\right)^2}\, \left[\mathrm{dilog}\left(1+\mathrm{e}^{-\beta eV_Q}\right)-
\mathrm{dilog}\left(1+\mathrm{e}^{\beta eV_Q}\right)\right], \label{dilog}
\end{eqnarray}
where $\mathrm{dilog}$ is the standard dilogarithm function \cite{Abramowitz}. One can finally use the definition of
differential capacitance per unit area, $C_Q=d\sigma/dV_Q$, to obtain from Eq.\ (\ref{dilog}) the quantum capacitance of a
single layer of graphene as \cite{Fang_2007}
\begin{eqnarray}
C_Q=\frac{2}{\lambda_Q}\,\ln\!\left[2\cosh\left(\beta eV_Q/2\right)\right], \label{CQ}
\end{eqnarray}
where we have defined the characteristic length scale for graphene,
\begin{eqnarray}
\lambda_Q=\frac{\pi}{2}\beta \left(\frac{\hbar v_F}{e}\right)^2, \label{lambdaQ}
\end{eqnarray}
with the value of $\lambda_Q\approx 18$ nm at RT. Note from Eq.\ (\ref{CQ}) that graphene's quantum capacitance grows
practically linearly with $V_Q$ when this potential exceeds the thermal potential, $V_\mathrm{th}=1/(e\beta)$, having the value
of $\approx$ 26 mV at RT.

We further assume that an upper surface of graphene is exposed to a thick layer of a symmetric $z\!\!:\!\!z$ electrolyte
containing the bulk number density per unit volume, $N$, for each kind of dissolved salt ions. Taking advantage of planar
symmetry, we place an $x$ axis perpendicular to graphene and pointing into the electrolyte. The theory developed by Borukhov
\emph{et al.}\cite{Borukhov_1997,Kilic_2007} to model finite ion size uses the mPB equation for the electrostatic potential
$V(x)$ in the electrolyte at a distance $x$ from graphene, given by
\begin{eqnarray}
\frac{d^2 V}{dx^2}=4\pi\frac{zeN}{\epsilon}\frac{2\sinh\left(\beta zeV\right)}{1+2\gamma\sinh^2\left(\beta zeV/2\right)},
 \label{mPB}
\end{eqnarray}
where $z(=1)$ is the valency of ions, $\epsilon$ is relative dielectric constant of water ($\approx 80$, assumed to be constant
throughout the electrolyte), and $\gamma=2a^3N$ is the packing parameter of the solvated ions, which are assumed to have same
effective size, equal to $a$ \cite{Borukhov_1997,Kilic_2007}. We note that the standard PB model is recovered from Eq.\
(\ref{mPB}) in the limit $\gamma\rightarrow 0$ \cite{Bard_2001}. By assuming the boundary condition $V(x)=0$ (and hence
$dV/dx=0$) at $x\rightarrow\infty$, deep into the electrolyte bulk, Eq.\ (\ref{mPB}) can be integrated once giving a relation
between the electric field and the potential at a distance $x$ from graphene. Assuming that graphene is placed at $x=0$, one can
use the boundary condition at the distance $d$ of closest approach for ions in the electrolyte to graphene,
\begin{eqnarray}
-\frac{dV}{dx}(d)=4\pi\frac{\sigma}{\epsilon},
 \label{boundary}
\end{eqnarray}
to establish a connection between the induced charge density on graphene, $\sigma$, and the potential drop, $V_D=V(d)$, across
the EDL as
\begin{eqnarray}
\sigma=\sqrt{\frac{2\epsilon N}{\pi\beta}}\sqrt{\frac{1}{2\gamma}\,\ln\!\left[1+2\gamma\sinh^2\left(\beta
zeV_D/2\right)\right]}\,\mathrm{sgn}(V_D).
 \label{EDL}
\end{eqnarray}

The total potential, $V_A$, applied between the reference electrode in the electrolyte and graphene can be written as
\begin{eqnarray}
V_A=V_\mathrm{pzc}+V_\mathrm{cl}+V_D+V_Q,
 \label{VA}
\end{eqnarray}
where $V_\mathrm{pzc}=\left(W_\mathrm{gr}-W_\mathrm{ref}\right)/e$ is the potential of zero charge \cite{Bard_2001} that stems
from difference between the work functions of graphene and the reference electrode, $W_\mathrm{gr}$ and $W_\mathrm{ref}$,
respectively, and $V_\mathrm{cl}=4\pi h\sigma/\epsilon^\prime$ is the potential drop across a charge free region between the
compact layer of the electrolyte ions condensed on the graphene surface, having the thickness $h$ on the order of the distance
of closest approach $d$ \cite{Kilic_2007,Kornyshev_2007}, and with $\epsilon^\prime<\epsilon$ taking into account a reduction of
the dielectric constant of water close to a charged wall \cite{Abrashkin_2007}. In our calculations, we shall neglect these two
contributions to the applied potential in Eq.\ (\ref{VA}) because $V_\mathrm{pzc}$ merely shifts the zero of that potential,
while a proper modeling of $V_\mathrm{cl}$ involves large uncertainty \cite{Kilic_2007,Kornyshev_2007}. However, usually the
effects of $V_\mathrm{cl}$ can be considered either small \cite{Kilic_2007} or incorporated in the mPB model via saturation of
the ion density at the electrolyte-graphene interface for high potential values \cite{Borukhov_1997}. Consequently, $V_D$ and
$V_Q$ represent the two main contributions in Eq.\ (\ref{VA}), with the $V_D$ being the surface potential of graphene that
shifts its Dirac point, and $V_Q$ being responsible for controlling the doping of graphene by changing its chemical potential.
Finally, we note that all results of our calculations will be symmetrical relative to the change in sign of the applied
potential because of our assumption that the effective sizes of the positive and negative ions are equal
\cite{Borukhov_1997,Kilic_2007}, but this constraint can be lifted by a relatively simple amendment to the mPB model
\cite{Kornyshev_2007}.

Using the relation $V_A=V_Q+V_D$ we obtain the total differential capacitance of the electrolytically top gated graphene as
\begin{eqnarray}
C^{-1}\equiv\frac{dV_A}{d\sigma}=\frac{dV_Q}{d\sigma}+\frac{dV_D}{d\sigma}=C_Q^{-1}+C_D^{-1},
\label{C}
\end{eqnarray}
where $C_Q(V_Q)$ is given in Eq.\ (\ref{CQ}), and $C_D(V_D)=d\sigma/dV_D$ is the differential capacitance per unit area of the
EDL, which can be obtained from Eq.\ (\ref{EDL}) as \cite{Kilic_2007,Kornyshev_2007},
\begin{widetext}
\begin{eqnarray}
C_D=\frac{\epsilon}{4\pi\lambda_D}\frac{\sinh\left(\beta ze\vert V_D\vert\right)}{\left[1+2\gamma\sinh^2\left(\beta
zeV_D/2\right)\right] \sqrt{\frac{2}{\gamma}\,\ln\!\left[1+2\gamma\sinh^2\left(\beta zeV_D/2\right)\right]}},
 \label{CD}
\end{eqnarray}
\end{widetext}
with $\lambda_D=\sqrt{\epsilon/(8\pi\beta z^2e^2N)}$ being the Debye length of the EDL \cite{Bard_2001}. Note that, in the limit
of a very low potential $V_D$, and hence for low density of ions at the graphene-EDL interface, one can set $\gamma\rightarrow
0$ in Eq.\ (\ref{CD}) to recover an expression for the EDL capacitance arising in the standard PB model \cite{Bard_2001},
\begin{eqnarray}
C_D\approx\frac{\epsilon}{4\pi\lambda_D}\,\cosh\left(\beta zeV_D/2\right).
 \label{CD_PB}
\end{eqnarray}
We further note that, while Eq.\ (\ref{CD_PB}) implies an unbounded growth of the EDL capacitance with $V_D$ in the PB model,
Eq.\ (\ref{CD}) suggests a non-monotonous behavior that will eventually give rise to a saturation of the total gate capacitance
at high applied voltages.

\section{Results}

Given the vast ranges of various parameters in our model, it is of interest to define reduced quantities. With the thermal
potential $V_\mathrm{th}=1/(e\beta)$, all potentials can be written as $\wtV=V/V_\mathrm{th}$. While typical regimes of graphene
doping require only $\vert\wtV_Q\vert\lesssim 50$, we shall extend this range in our calculations up to about
$\vert\wtV_Q\vert\approx 100$ to represent doping levels attained in recent experiments in electrolytic environment
\cite{Das_2008}. Next, referring to Eq.\ (\ref{dilog}), we define the characteristic number density of doped charge carriers in
graphene by $n_0=(2/\pi)/\left(\beta\hbar v_F\right)^2$, which has the value of $n_0\approx 10^{11}$ cm$^{-2}$ at RT. Therefore,
defining the reduced density by $\wtn=n/n_0$, and hence $\wtsig=\sigma/(en_0)=\wtn$, we note that $\vert\wtn\vert$ may reach up
to around 10$^3$ \cite{Castro_2009,Das_2008}. It is worthwhile mentioning that graphene's characteristic parameters $\lambda_Q$
and $n_0$ are related via $\epsilon\lambda_B\lambda_Qn_0=1$, where $\lambda_B=\beta e^2/\epsilon$ is the Bjerrum length of the
aqueous environment, taking the value of $\lambda_B\approx$ 0.7 nm at RT \cite{Bard_2001}. Furthermore, it follows from Eq.\
(\ref{CQ}) that the natural unit of capacitance for this system is $C_0=en_0/V_\mathrm{th}=\lambda_Q^{-1}$, taking the value of
$C_0\approx 0.6$ $\mu$F/cm$^2$ at RT. Turning now to Eq.\ (\ref{EDL}), one can define the characteristic number density of ions
per unit volume in the solution by
\begin{eqnarray}
N_0=\frac{\pi}{2}n_0^2\lambda_B=\frac{2e^2}{\pi\epsilon\beta^3(\hbar v_F)^4}, \label{N0}
\end{eqnarray}
which takes the value of $N_0\approx 1.08\times 10^{-6}$ nm$^{-3}$ $\approx$ 1.8 $\mu$M at RT. Defining the reduced
concentration of ions in the bulk of the electrolyte by $\wtN=N/N_0$, it would be of interest to explore a broad range of its
values, e.g., $10^{-1}<\wtN<10^5$. Finally, in order to estimate the packing parameter, we write $\gamma=\nu\wtN$ and take
$a=\lambda_B$ to obtain $\nu=2\lambda_B^3 N_0\approx 7\times 10^{-7}$. With these definitions, Eqs.\ (\ref{dilog})
and(\ref{EDL}) now read, respectively,
\begin{eqnarray}
\wtsig=\mathrm{dilog}\left(1+\mathrm{e}^{-\wtV_Q}\right)- \mathrm{dilog}\left(1+\mathrm{e}^{\wtV_Q}\right), \label{wtdilog}
\end{eqnarray}
\begin{eqnarray}
\wtsig=\sqrt{\frac{1}{2\nu}\,\ln\!\left[1+2\nu\wtN\sinh^2\left(z\wtV_D/2\right)\right]}\,\mathrm{sgn}(\wtV_D).
 \label{wtEDL}
\end{eqnarray}
\begin{figure}
\begin{center}
\includegraphics[width=90mm]{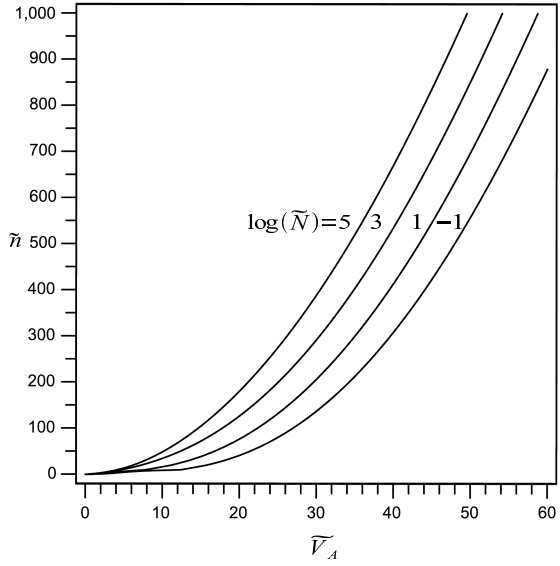}
\vspace{-22mm} \caption{Reduced density $\wtn$ of doped charge carriers in graphene versus the reduced applied voltage $\wtV_A$
for several values of the reduced salt concentration $\wtN$ in a NaF aqueous solution.}
\end{center}
\end{figure}
\begin{figure}
\begin{center}
\includegraphics[width=90mm]{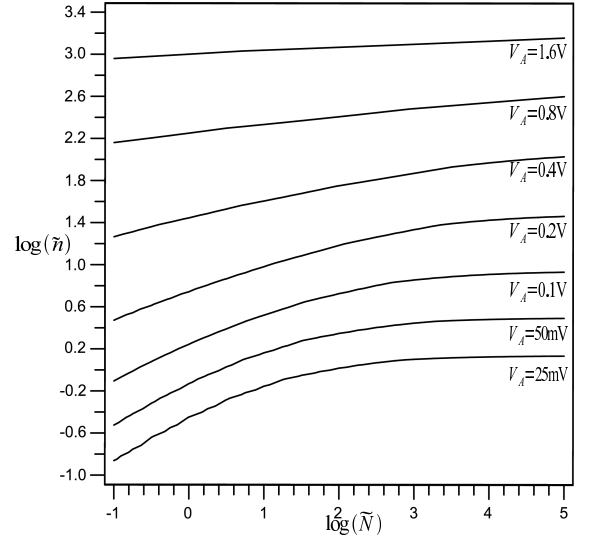}
\vspace{-22mm} \caption{Reduced density $\log_{10}(\wtn)$ of doped charge carriers in graphene versus the reduced salt
concentration $\log_{10}(\wtN)$ for several values of the applied voltage $V_A$ (in Volts) for a NaF aqueous solution.}
\end{center}
\end{figure}

We now use Eqs.\ (\ref{wtdilog}) and (\ref{wtEDL}) in conjunction with the relation $\wtV_A=\wtV_Q+\wtV_D$ to eliminate the
potential components $\wtV_Q$ and $\wtV_D$, and to evaluate the reduced density of doped charge carriers in graphene,
$\wtn=\wtsig$, as a function of the reduced applied voltage $\wtV_A$ and the reduced salt concentration $\wtN$. The results are
shown in Figs.\ 1 and 2, covering the following ranges: $\vert\wtn\vert\leq 10^3$ (corresponding to $\vert n\vert \lesssim
10^{14}$ cm$^{-2}$), $\vert\wtV_A\vert \leq 60$ (corresponding to $\vert V_A\vert\lesssim$ 1.6 V), and $10^{-1}\leq\wtN\leq
10^5$ (corresponding to $0.18$ $\mu$M $ \lesssim N \lesssim$ 0.18 M). In Fig.\ 1 one notices a strong dependence of $\wtn$ on
the applied potential for $\wtV_A$ greater than about 30, which gives approximately equal rates of change for each salt
concentration at the highest values of the applied potential. On the other hand, at the lower applied potential values, there
exists a much stronger dependence on the salt concentration, which is revealed in Fig.\ 2, showing $\log_{10}\wtn$ versus
$\log_{10}\wtN$ for several applied voltages. Indeed, one notices a very strong sensitivity of the doped charge carrier density
in graphene to the salt concentration for applied voltages $V_A\lesssim$ 0.4 V in the range of salt concentrations $N\lesssim$ 1
mM. Even though this sensitivity seems to be the strongest at the lowest applied voltages, one should bear in mind that the
electrical conductivity in graphene becomes rather uncertain around its minimum value, which extends up to doping densities
about $n\approx 10^{11}$ cm$^{-2}$ \cite{Tan_2007,Adam_2009}. Therefore, it seems that $0.1\lesssim V_A\lesssim 0.4$ V would be
an optimal range of applied voltages for sensor applications of the electrolytically top gated graphene in probing salt
concentrations in the sub-microMole range.

Next, moving to the capacitance of electrolytically top gated graphene, $C$, we note that the reduced capacitances,
$\wtC_Q=C_Q/C_0$ and $\wtC_D=C_D/C_0$, are obtained from Eqs.\ (\ref{CQ}) and (\ref{CD}) as
\begin{eqnarray}
\wtC_Q=2\,\ln\!\left[2\cosh\left(\wtV_Q/2\right)\right], \label{wtCQ}
\end{eqnarray}
\begin{widetext}
\begin{eqnarray}
\wtC_D=\frac{z}{2}\frac{\wtN\,\sinh\left(z\vert\wtV_D\vert\right)}{\left[1+2\nu\wtN\sinh^2\left(z\wtV_D/2\right)\right]
\sqrt{\frac{2}{\nu}\,\ln\!\left[1+2\nu\wtN\sinh^2\left(z\wtV_D/2\right)\right]}},
 \label{wtCD}
\end{eqnarray}
\end{widetext}
showing that $\wtC_Q$ and $\wtC_D$ are comparable in magnitude for vanishing potentials when the salt concentration is $\wtN\sim
1$. Moreover, referring to Eq.\ (\ref{C}) as an electrical model where graphene and the EDL act as a series connection of
capacitors, it follows that graphene's quantum capacitance $C_Q$ will be promoted as the dominant contribution to the total gate
capacitance as the salt concentration increases.
\begin{figure}
\begin{center}
\includegraphics[width=85mm]{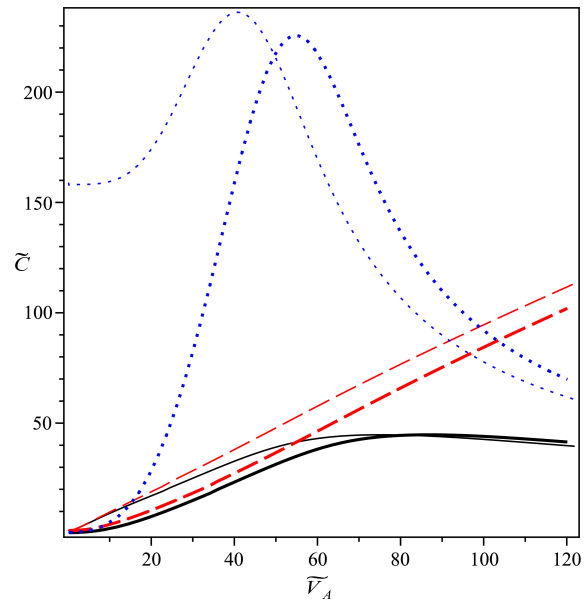}
\vspace{-22mm} \caption{The dependence on the reduced applied voltage $\wtV_A$ is shown for: the total reduced capacitance
$\wtC=\wtC_Q\wtC_D/\left(\wtC_Q+\wtC_D\right)$ (solid black lines), graphene's reduced quantum capacitance $\wtC_Q$ (dashed red
lines), and the reduced capacitance of the electric double layer $\wtC_D$ (dotted blue lines), in the NaF aqueous solutions with
the reduced salt concentrations of $\wtN$ = 1 (thick lines) and $10^5$ (thin lines).}
\end{center}
\end{figure}

We now use the equality of the right-hand-sides in Eqs.\ (\ref{wtdilog}) and (\ref{wtEDL}) along with the relation
$\wtV_A=\wtV_Q+\wtV_D$ to eliminate $\wtV_Q$ and $\wtV_D$, and to evaluate the reduced quantum capacitance of graphene from Eq.\
(\ref{wtCQ}), as well as the reduced capacitance of the EDL from Eq.\ (\ref{wtCD}) as functions of the reduced applied voltage
$\wtV_A$. Results are shown in Fig.\ 3 along with the total reduced capacitance of the system based on Eq.\ (\ref{C}), for two
reduced salt concentrations, $\wtN$ = 1 and $10^5$ (corresponding to $N\approx$ 1.8 $\mu$M and 0.18 M, respectively). We show
our results for the reduced applied voltages up to $\wtV_A\approx$ 120 in order to elucidate the effect of saturation in the
total capacitance that occurs at $\wtV_A\approx$ 85 V (corresponding to $V_A\approx$ 2.21 V) for $\wtN=1$ and at $\wtV_A\approx$
75 (corresponding to $V_A\approx$ 1.95 V) for $\wtN=10^5$. As can be seen from dotted curves in Fig.\ 3, showing a
non-monotonous dependence of the EDL capacitance on the applied voltage, the saturation effect in the total capacitance of the
electrolytically top gated graphene is a consequence of the steric effect of the electrolyte ions that are crowded at the
graphene surface at high applied voltages \cite{Kilic_2007}. Even though the voltages where the saturation occurs are relatively
high, they may still be accessible in experiments on graphene. Furthermore, we see in Fig.\ 3 that, at intermediate applied
voltages, the rate of change of the total capacitance follows closely that of the quantum capacitance, with the value $\approx$
23 $\mu$F/(V\,cm$^2$) that is commensurate with recent measurement \cite{Xia_2009}. At the lowest applied voltages, one notices
in Fig.\ 3 a ``rounding'' of the total capacitance as a function of voltage for low salt concentrations, which comes from the
EDL capacitance. Such rounding is observed in the recent experiment \cite{Xia_2009}.

\begin{figure}
\begin{center}
\includegraphics[width=85mm]{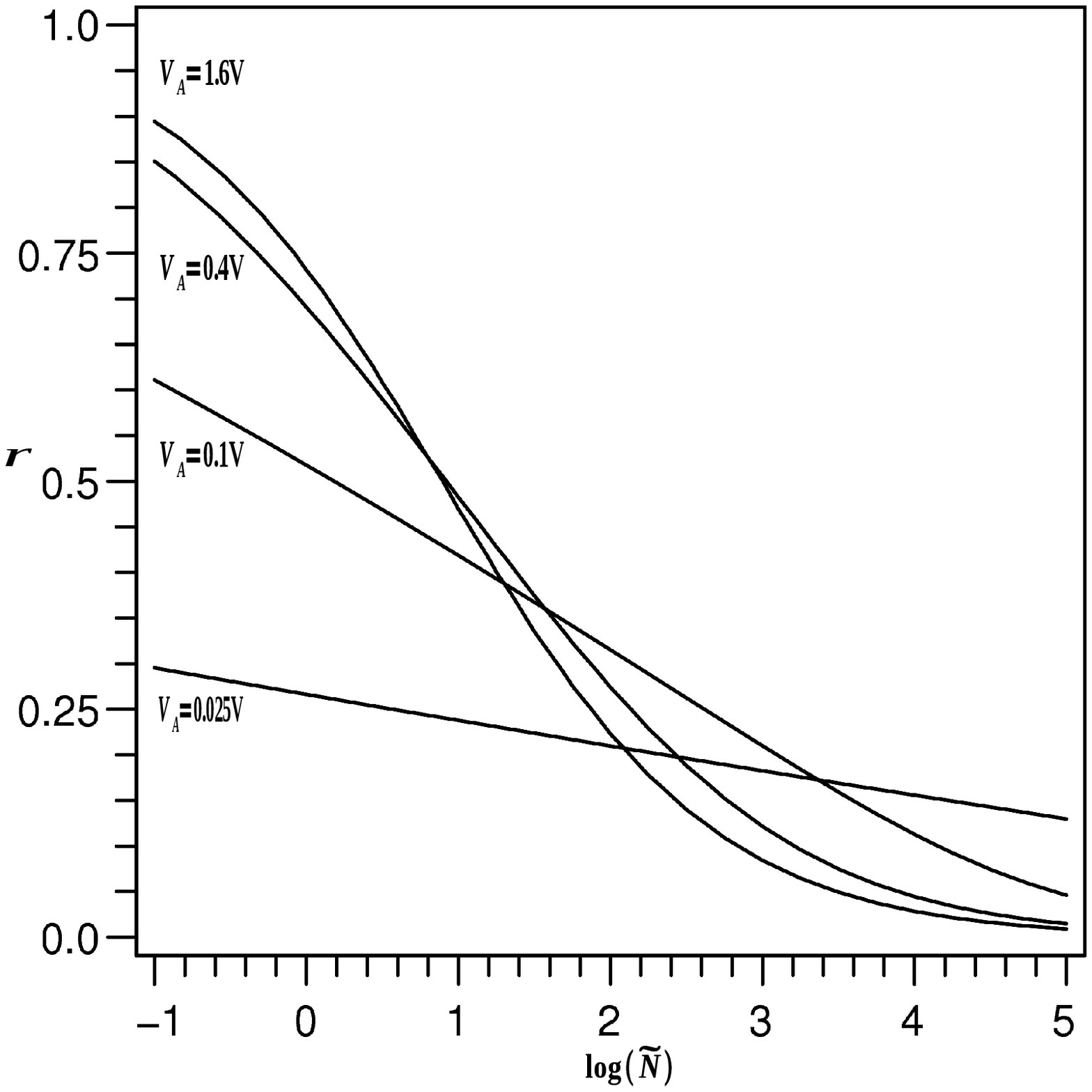}
\caption{The dependence of the ratio $r=V_D/V_A$ on the reduced salt concentration $\log_{10}(\wtN)$ for several values of the
applied voltage $V_A$ (in Volts) for a NaF aqueous solution.}
\end{center}
\end{figure}
As a consequence of the vast differences between the capacitances shown in Fig.\ 3, one expects that there exists a broad
variation in the way how is the total applied voltage $V_A$ split between the potential drop $V_D$ across the EDL and the
voltage $V_Q$ pertaining to the change in graphene's chemical potential. We therefore display in Fig.\ 4 the variation of the
fraction $r=V_D/V_A$ as a function of the reduced salt concentration in the electrolyte $\wtN$ for several values of the applied
voltage $V_A$. One can see that, at low salt concentrations, the potentials $V_Q$ and $V_D$ are roughly comparable in magnitude,
although the ratio $r$ increases in favor of the potential drop across the electrolyte as the applied voltage increases.
However, this trend is reversed at high salt concentrations and, more importantly, Fig.\ 4 shows that most of the applied
voltage is used to increase graphene's chemical potential for a full range of applied voltages when salt concentration $N$
exceeds, say, mM. Besides its importance for applications, the fact that the potential drop across the electrolyte remains very
small at high applied voltages also alleviates concern that a high electric filed in the electrolyte may cause the onset of
voltage-dependent electrochemical reactions on graphene.

\section{Concluding remarks}

We have analyzed the doping of single-layer graphene due to application of the gate potential through an aqueous solution of
salt using a modified Poisson-Boltzman model for electrolyte, and found great sensitivity of the induced charge density in
graphene to the broad ranges of both salt concentration and applied voltage. We have further analyzed differential capacitance
of the electrolytically top gated graphene, and found that its quantum capacitance is promoted as the dominant component owing
to a reduction in the Debye length of the electric double layer when the salt concentration increases. In this case, very little
potential drop appears across the electrolyte, and graphene takes most of the voltage drop to shift its chemical potential.
These findings have several important consequences.

First, since graphene's conductivity is dependent upon its chemical potential via doping density, Eq.\ (\ref{density}), the
sensitivity of the latter to the salt concentration implies good prospects for applications in biochemical sensors, especially
for in-vivo electrochemical measurements in biological systems owing to graphene's natural bio-compatibility. Next, since most
of the applied voltage can be used to increase the chemical potential of graphene, as opposed to a potential drop across the
electrolyte, one can envision ways to use a very thin top gate (in the form of a liquid or solid electrolyte) that requires
relatively low gate voltage to change the chemical potential (and hence conductivity) of graphene in future small scale field
effect devices with tunable conductivity. Among other aspects of the increased role of graphene's quantum capacitance is
reduction of the electrical field across the electrolyte. This can help reduce the rates of voltage-induced electrochemical
reactions on graphene's surface, as well as improve the mobility of charge carriers in graphene by reducing their scattering
rates on various impurities. Moreover, since quantum capacitance is basically the capacitance associated with change in carrier
densities in graphene, it can be seen as analogous to the junction capacitance, and the smaller quantum capacitance could in
turn lead to faster switching time for graphene based devices.

Many of these advantages of top gating through an electrolyte are related to a high bulk dielectric constant of the electrolyte,
especially in aqueous solutions. So, even though the oxide thickness can be reduced down to around 2nm in the present generation
conventional MOS structures, the much higher dielectric constant of water in comparison to SiO$_2$ should provide for a higher
gate capacitance, translating into much better field effect performance, as discussed above. However, dielectric constant of an
electrolyte can be significantly reduced close to a charged surface \cite{Kornyshev_2007,Abrashkin_2007}, and this issue has yet
to be discussed in the context of electrolytic top gating of carbon nano-structures.

\begin{acknowledgments}
This work was supported by the Natural Sciences and Engineering Research Council of Canada.
\end{acknowledgments}

\bibliography{review}{}

\end{document}